# The complexity of conservative finite-valued CSPs

Vladimir Kolmogorov[*]
University College London
v.kolmogorov@cs.ucl.ac.uk

Stanislav Živný[†]
Oxford University
standa.zivny@comlab.ox.ac.uk

**Abstract**

We study the complexity of valued constraint satisfaction problems (VCSP). A problem from VCSP is characterised by a *constraint language*, a fixed set of cost functions over a finite domain. An instance of the problem is specified by a sum of cost functions from the language and the goal is to minimise the sum. We consider the case of so-called *conservative* languages; that is, languages containing all unary cost functions, thus allowing arbitrary restrictions on the domains of the variables. This problem has been studied by Bulatov [LICS'03] for $\{0, \infty\}$-valued languages (i.e. CSP), by Cohen *et al.* (AIJ'06) for Boolean domains, by Deineko *et al.* (JACM'08) for $\{0, 1\}$-valued cost functions (i.e. Max-CSP), and by Takhanov (STACS'10) for $\{0, \infty\}$-valued languages containing all finite-valued unary cost functions (i.e. Min-Cost-Hom).

We give an elementary proof of a complete complexity classification of conservative finite-valued languages: we show that every conservative finite-valued language is either tractable or NP-hard. This is the *first* dichotomy result for finite-valued VCSPs over non-Boolean domains.

[*]Vladimir Kolmogorov is supported by the Royal Academy of Engineering/EPSRC.

[†]Stanislav Živný is supported by Junior Research Fellowship at University College, Oxford. Part of this work was done while the second author was visiting Microsoft Research Cambridge.

# 1 Introduction

The constraint satisfaction problem is a central generic problem in computer science. It provides a common framework for many theoretical problems as well as for many real-life applications, see [1] for a nice survey. An instance of the *constraint satisfaction problem* (CSP) consists of a collection of variables which must be assigned values subject to specified constraints. CSP is known to be equivalent the problem of evaluating conjunctive queries on databases [2], and to the homomorphism problem for relations structures [3].

An important line of research on the CSP is to identify all tractable cases; that is, cases that are recognisable and solvable in polynomial time. Most of this work has been focused on one of the two general approaches: either identifying structural properties of the way constrains interact which ensure tractability no matter what forms of constraint are imposed [4], or else identifying forms of constraint which are sufficiently restrictive to ensure tractability no matter how they are combined [5, 3].

The first approach has been used to characterise all tractable cases of bounded-arity CSPs: the *only* class of structures which ensures tractability (subject to a certain complexity theory assumption, namely FPT $\neq$ W$[1]$) are structures of bounded tree-width modulo homomorphic equivalence [6, 7]. The second approach has led to identifying certain algebraic properties known as polymorphisms [8] which are necessary for a set of constraint types to ensure tractability. A set of constraint types which ensures tractability is called a *tractable constraint language*.

Schaefer in his seminal work [9] gave a complete complexity classification of Boolean constraint languages. The algebraic approach based on polymorphisms [10] has been so far the most successful tool in generalising Schaefer's result to languages over a 3-element domain [11], conservative languages [12], languages comprising of a single binary relation without sources and sinks [13] (see also [14]), and languages comprising of a single binary relation that is a special triad [15]. The algebraic approach has also been essential in characterising the power of local consistency [16] and the "few subpowers property" [17], the two main tools known for solving tractable CSPs. The ultimate goal in this line of research is to answer the *Dichotomy Conjecture* of Feder and Vardi, which states that every constraint language is either tractable or NP-hard [3]. We remark that there are other approaches to the dichotomy conjecture; see, for instance, [1] for a nice survey of Hell and Nešetřil, and [18] for a connection between the Dichotomy Conjecture and probabilistically checkable proofs.

Since in practice many constraint satisfaction problems are over-constrained, and hence have no solution, or are under-constrained, and hence have many solutions, *soft* constraint satisfaction problems have been studied [19]. In an instance of the soft CSP, every constraint is associated with a function (rather than a relation as in the CSP) which represents preferences among different partial assignments, and the goal is to find the best assignment. Several very general soft CSP frameworks have been proposed in the literature [20, 21]. In this paper we focus on one of the very general frameworks, the *valued* constraint satisfaction problem (VCSP) [20].

Similarly to the CSP, an important line of research on the VCSP is to identify tractable cases which are recognisable in polynomial time. Is is well known that structural reasons for tractability generalise to the VCSP [19]. In the case of language restrictions, only a few conditions are known to guarantee tractability of a given set of valued constraints [22, 23].

**Related work** Creignou *et al.* have studied various generalisations of the CSP to optimisation problems over Boolean domains [24], see also [25, 26]. Cohen *et al.* have completely classified the complexity of valued languages over Boolean domains [22].

Bulatov, making use of the techniques of universal algebra, has completely classified the complexity of conservative $\{0,\infty\}$-valued languages, i.e. languages containing all possible $\{0,\infty\}$-valued unary cost functions. (Note that $\{0,\infty\}$-valued languages are just normal constraint languages, that is, sets of relations.) Using the algebraic approach, Takhanov has completely classified the complexity of $\{0,\infty\}$-valued languages containing additionally all finite-valued unary cost functions [27]. (Note that this problem is equivalent to the *minimum-cost homomorphism* problem.) It is not clear how to use the algebraic methods of Bulatov and Takhanov for our problem as the algebraic methods developed so far work only for $\{0,\infty\}$-valued VCSPs, that is, CSPs. (One possible way might be to use results of Cohen *et al.* [28], but our results and techniques do not rely on results in [28].)

Deineko *et al.* [29] have completely classified the complexity of conservative $\{0,1\}$-valued cost function languages.[1] (Note that with respect to polynomial-time solvability $\{0,1\}$-valued VCSPs and Max-CSP are equivalent.) The crucial result in [29] is to show that if a conservative $\{0,1\}$-valued language $\Gamma$ is not submodular with respect to any order, then $\Gamma$ can express (without changing the complexity) language $L'$ that is not submodular with respect to any order and $L'$ consists of at most 3 binary cost functions over a domain of size at most $4$. In the $\{0,1\}$-valued case, there are only finitely many of these cost functions, and a cleverly-pruned computer search had shown that all such languages are intractable [29]. It is not clear how to use the techniques from [29] for the general VCSP framework with arbitrary costs.

Raghavendra [30] and Raghavendra and Steurer [31] have shown how to optimally approximate any finite-valued VCSP.

**Contributions** We prove a dichotomy theorem for finite-valued conservative languages: we show that a conservative finite-valued language $\Gamma$ is tractable if and only if $\Gamma$ admits a symmetric tournament pair multimorphism [23]. This result is known for Boolean conservative finite-valued languages [22]. The Boolean case has been resolved using the notion of *multimorphisms* [22]. We will also use multimorphisms here to derive the first complete complexity classification over domains of arbitrary size. To compare our result with [29], we show that even in the more general framework of finite-valued languages (as opposed to $\{0,1\}$-valued languages), there is *only* one reason for tractability, which is the same as in [29].[2]

Despite the fact that we use the notion of multimorphisms, our proofs are elementary and largely self-contained. We do not rely on the heavy algebraic machinery that has been used in the complexity classification of conservative $\{0,\infty\}$-valued languages by Bulatov [12], and in the complexity classification of $\{0,\infty\}$-valued languages containing additionally all finite-valued unary cost functions by Takhanov [27].

Naturally, our approach is similar to the one of Bulatov and Takhanov: the complexity of Boolean conservative languages is known, and hence given a non-Boolean language, we explore the interactions between different 2-element subdomains. Given a conservative finite-valued language $\Gamma$, we will investigate properties of a certain graph $G_\Gamma$ associated with the language and cost functions expressible over $\Gamma$. The graph is finite, but not easy to construct. We show that $\Gamma$ is tractable if and only if $G_\Gamma$ satisfies certain properties. Thus we obtain a dichotomy theorem. However, in order to test whether a given $\Gamma$ is tractable, we do not need to construct $G_\Gamma$ as it follows from our result that we just need to test for the existence of a certain multimorphism of $\Gamma$.

Even though our results are formulated as valued constraint satisfaction problems, they apply to various other optimisation frameworks that are equivalent to valued constraint problems such as Gibbs energy minimisation, Markov Random Fields and other graphical models [32, 33].

**Organisation of the paper** The rest of the paper is organised as follows. In Section 2, we define valued constraint satisfaction problems (VCSPs), conservative languages, multimorphisms and other necessary definitions needed throughout the paper. We state our results in Section 3 and prove them in Section 4.

[1] In fact, results in [29] are stronger in two ways. Firstly, Deineko *et al.* have classified the complexity of languages containing the so-called fixed-value constraints; that is, constraints of the form "x=d", where $x$ is a variable and $a$ is a domain value. Secondly, Deineko *et al.* have considered the weighted version of the problem, which allows the range of every cost function $f$ to be $\{0, c_f\}$, where $c_f$ depends on $f$. (The $\{0,1\}$-valued case is the special case when $c_f = 1$ for every cost function $f$.)

[2] It follows from [23] that in the finite-valued case, having a symmetric tournament pair multimorphism is the same as being submodular with respect to some order.

Finally, in Section 5 we discuss the general case, which is still open. We review some known results and state our conjecture for the general case.

## 2 Background and notation

We denote by $\mathbb{R}_+$ the set of all non-negative real numbers. We denote $\overline{\mathbb{R}}_+ = \mathbb{R}_+ \cup \{\infty\}$ with the standard addition operation extended so that for all $a \in \mathbb{R}_+$, $a + \infty = \infty$. Members of $\overline{\mathbb{R}}_+$ are called *costs*.

Throughout the paper, we denote by $D$ any fixed finite set, called a *domain*. Elements of $D$ are called *domain values* or *labels*. For distinct labels $a, b, \ldots$ in $D$ we denote $D_a = \{a\}$, $D_{ab} = \{a, b\}$ and so on.

A function $f$ from $D^m$ to $\overline{\mathbb{R}}_+$ will be called a *cost function* on $D$ of *arity* $m$. If the range of $f$ lies entirely within $\mathbb{R}$, then $f$ is called a *finite-valued* cost function. If the range of $f$ is $\{0, \infty\}$, then $f$ is called a *crisp* cost function. If the range of a cost function $f$ includes non-zero finite costs and infinity, we emphasise this fact by calling $f$ a *general-valued* cost function. Let $f : D^m \to \overline{\mathbb{R}}_+$ be an $m$-ary cost function $f$. We denote $\texttt{dom} f = \{\boldsymbol{x} \in D^m \mid f(\boldsymbol{x}) < \infty\}$ to be the effective domain of $f$. Functions $f$ of arity $m = 2$ are called *binary*.

A *language* is a set of cost functions with the same set $D$. Language $\Gamma$ is called finite-valued (crisp, general-valued respectively) if all cost functions in $\Gamma$ are finite-valued (crisp, general-valued respectively). A language $\Gamma$ is *Boolean* if $|D| = 2$.

**Definition 1.** *An instance $\mathcal{I}$ of the* valued constraint satisfaction problem *(VCSP) is a function $D^V \to \overline{\mathbb{R}}_+$ given by*

$$Cost_{\mathcal{I}}(\boldsymbol{x}) = \sum_{t \in T} f_t\left(x_{i(t,1)}, \ldots, x_{i(t,m_t)}\right)$$

*It is specified by a finite set of nodes $V$, finite set of terms $T$, cost functions $f_t : D^{m_t} \to \overline{\mathbb{R}}_+$ or arity $m_t$ and indices $i(t,k) \in V$ for $t \in T$, $k = 1, \ldots, m_t$. A* solution *to $\mathcal{I}$ is an assignment $\boldsymbol{x} \in D^V$ with the minimum cost.*

We denote by $\mathsf{VCSP}(\Gamma)$ the class of all VCSP instances whose terms $f_t$ belong to $\Gamma$. A finite language $\Gamma$ is called *tractable* if $\mathsf{VCSP}(\Gamma)$ can be solved in polynomial time, and *intractable* if $\mathsf{VCSP}(\Gamma)$ is NP-hard. An infinite language $\Gamma$ is tractable if every finite subset $\Gamma' \subseteq \Gamma$ is tractable, and intractable if there is a finite subset $\Gamma' \subseteq \Gamma$ that is intractable.

Intuitively, a language is conservative if one can restrict the domain of any variable to an arbitrary subset of the domain.

**Definition 2.** *Language $\Gamma$ is called* conservative *if $\Gamma$ contains all finite-valued unary cost functions $u : D \to \mathbb{R}_+$.*

We remark that in the crisp case, conservative languages are defined differently in the literature: a crisp language is called conservative if it contains all possible $\{0, \infty\}$-valued unary cost functions [12]. In this paper we always use Definition 2, unless explicitly referring to the crisp case.

**Definition 3.** *Let $\langle \sqcap, \sqcup \rangle$ be a pair of operations, where $\sqcap, \sqcup : D \times D \to D$.*

- *Pair $\langle \sqcap, \sqcup \rangle$ is called a* multimorphism *for function $f : D^m \to \overline{\mathbb{R}}_+$ if*

$$f(\boldsymbol{x} \sqcap \boldsymbol{y}) + f(\boldsymbol{x} \sqcup \boldsymbol{y}) \leq f(\boldsymbol{x}) + f(\boldsymbol{y}) \qquad \forall \boldsymbol{x}, \boldsymbol{y} \in \texttt{dom} f \tag{1}$$

  *where operations $\sqcap, \sqcup$ are applied component-wise.*

- *Pair $\langle \sqcap, \sqcup \rangle$ is called* conservative *if $\{a \sqcap b, a \sqcup b\} = \{a, b\}$ for all $a, b \in D$.*

- *Pair* $\langle \sqcap, \sqcup \rangle$ *is called a* symmetric tournament pair (STP) *if it is conservative and both operations* $\sqcap, \sqcup$ *are commutative, i.e.* $a \sqcap b = b \sqcap a$ *and* $a \sqcup b = b \sqcup a$ *for all* $a, b \in D$.

We say that $\langle \sqcap, \sqcup \rangle$ is a multimorphism of language $\Gamma$, or $\Gamma$ admits $\langle \sqcap, \sqcup \rangle$, if all cost functions $f \in \Gamma$ satisfy (1). Using a polynomial-time algorithm for minimising submodular functions, Cohen *et al.* have obtained the following result:

**Theorem 4** ([23])**.** *If a language* $\Gamma$ *admits an STP, then* $\Gamma$ *is tractable.*

Cohen *et al.* have completely classified the tractability of Boolean languages, i.e. the case $|D| = 2$. The following is a simple corollary of their classification:

**Theorem 5** ([22])**.** *Let* $\Gamma$ *be a Boolean conservative finite-valued language. If* $\Gamma$ *admits an STP then it is tractable; otherwise it is NP-hard.*

The main result of this paper is a generalisation of this result to non-Boolean domains.

Finally, we define the important notion of expressibility, which captures the idea of introducing auxiliary variables in a VCSP instance and the possibility of minimising over these auxiliary variables. (For crisp languages, this is equivalent to *implementation* [25].)

**Definition 6.** *A cost function* $f : D^m \to \overline{\mathbb{R}}_+$ *is* expressible *over a language* $\Gamma$ *if there exists an instance* $\mathcal{I} \in \mathsf{VCSP}(\Gamma)$ *with the set of nodes* $V = \{1, \ldots, m, m+1, \ldots, m+k\}$ *where* $k \geq 0$ *such that*

$$f(\boldsymbol{x}) = \min_{\boldsymbol{y} \in D^k} Cost_{\mathcal{I}}(\boldsymbol{x}, \boldsymbol{y}) \qquad \forall \boldsymbol{x} \in D^m$$

*We define* $\Gamma^*$ *to be the* expressive power *of* $\Gamma$*; that is, the set of all cost functions* $f$ *such that* $f$ *is expressible over* $\Gamma$.

The importance of expressibility is in the following result:

**Theorem 7** ([22])**.** *For any language* $\Gamma$*,* $\Gamma$ *is tractable iff* $\Gamma^*$ *is tractable.*

It is easy to observe and well known that if $\langle \sqcap, \sqcup \rangle$ is a multimorphism of $\Gamma$, then it is also a multimorphism of $\Gamma^*$ [22].

From now on we assume that $\Gamma$ is a fixed conservative language.

## 3 Our results

In this section, we relate the complexity of a conservative language $\Gamma$ to properties of a certain graph $G_\Gamma$ associated with $\Gamma$.

Consider undirected graph $G_\Gamma = (P, E)$ where the set of nodes is $P = \{(a, b) \mid a, b \in D, a \neq b\}$ and the set of edges $E$ is defined as follows: there is an edge between $(a, b) \in P$ and $(a', b') \in P$ iff there exists binary cost function $f \in \Gamma^*$ such that

$$f(a, a') + f(b, b') > f(a, b') + f(b, a') \;, \quad (a, b'), (b, a') \in \mathtt{dom} f \qquad (2)$$

Note that $G_\Gamma$ may have self-loops. For node $p \in P$ we denote the self-loop by $\{p, p\}$. We say that edge $\{(a, b), (a', b')\} \in E$ is *soft* if there exists binary $f \in \Gamma^*$ satisfying (2) such that at least one of the assignments $(a, a')$, $(b, b')$ is in $\mathtt{dom} f$. Edges in $E$ that are not soft are called *hard*. For node $p = (a, b) \in P$ we denote $\bar{p} = (b, a) \in P$. Note, a somewhat similar graph (but not the same) was used by Takhanov [27] for languages $\Gamma$ containing crisp functions and finite unary terms.[3]

We denote $M \subseteq P$ to be the set of vertices $(a, b) \in P$ without self-loops, and $\overline{M} = P - M$ to be the complement of $M$. It is not difficult to show that $(a, b) \in M$ iff $(b, a) \in M$ (see next section).

**Definition 8.** *A conservative pair of operations $\langle \sqcap, \sqcup \rangle : D \times D \to D$ is called an* STP on $M$ *if it is commutative on $M$, i.e.* $a \sqcap b = b \sqcap a$ *and* $a \sqcup b = b \sqcup a$ *for all* $(a, b) \in M$.

Our main result is the following

**Theorem 9.**

*(a) If $E$ has a soft self-loop $\{(a, b), (a, b)\}$, then $\Gamma$ is NP-hard.*

*(b) If $E$ does not have soft self-loops, then $\Gamma$ admits an STP on $M$.*

A proof is given in the next section. When combined with the result of Cohen *et al.* [23], Theorem 9 gives a dichotomy result for conservative finite-valued languages:

**Corollary 10.** *Let $\Gamma$ be a conservative finite-valued language. If $\Gamma$ admits an STP, then $\Gamma$ is tractable; otherwise $\Gamma$ is NP-hard.*

*Proof.* Consider the graph $G_\Gamma$ associated with $\Gamma$. If $G_\Gamma$ contains a soft self-loop, then, by Theorem 9(a), $\Gamma$ is intractable. Suppose that $G_\Gamma$ does not contain soft self-loops. As $\Gamma$ is finite-valued, $G_\Gamma$ cannot have hard self-loops. Therefore, $\overline{M}$ is empty and $M = P$. By Theorem 9(b), $\Gamma$ admits an STP. The tractability then follows from Theorem 4. □

**Remark 1** Note that we can determine whether $\Gamma$ is tractable by testing the existence of an STP of $\Gamma$. For a fixed finite $D$ and $\Gamma$, there are only finitely many potential STPs. We do not need to build the graph $G_\Gamma$, whose properties depend on $\Gamma^*$.

**Remark 2** Our results can be easily extended to cost functions with the range $\mathbb{R}$; that is, allowing finite negative costs. This is due to the fact that for any finite language $\Gamma$, we can add a finite cost to all cost functions from $\Gamma$ to make the costs non-negative and this does not change the complexity of $\Gamma$.

**Remark 3** Let $u_d(x) = 0$ if $x = d$ and $u_d(x) = c$ if $x \neq d$, where $c \in \mathbb{R}_+$ is a fixed non-zero cost. Cost functions $u_d$, where $d \in D$, are called *fixed-value* cost functions. It is easy to see that our results classify the complexity of fixed-value languages. Indeed, as fixed-value languages are a subset of conservative languages, the tractability result remains the same, and the finite-value case of Theorem 9(a) remains valid as well. However, this is not the case for crisp languages where a complexity classification of crisp languages including crisp fixed-value cost functions would imply a complexity classification of all crisp languages [5], where crisp fixed-value cost functions are defined as $u_d(x) = 0$ if $x = d$ and $u_d(x) = \infty$ if $x \neq d$.

## 4 Proof of Theorem 9

In Section 4.1 we will first prove part (a). Then in Section 4.2 we will prove some properties of $G_\Gamma$ assuming that $G_\Gamma$ does not have self-loops. Using these properties, we will construct an STP on $M$ in Section 4.3.

[3] Roughly speaking, the graph structure in [27] was defined via a "min" polymorphism rather than a $\langle \min, \max \rangle$ multimorphism, so the property $\{p, q\} \in E \Rightarrow \{\bar{p}, \bar{q}\} \in E$ (that we prove for our graph in the next section) might not hold in Takhanov's case. Also, in [27] edges were not classified as being soft or hard.

### 4.1 NP-hard case

In this section we prove Theorem 9(a). From the assumption, there is a binary $f \in \Gamma^*$ such that $f(a,a) + f(b,b) > f(a,b) + f(b,a)$, and at least one of the assignments $(a,a),(b,b)$ is in $\mathrm{dom} f$. First, let us assume that both $(a,a)$ and $(b,b)$ are in $\mathrm{dom} f$. Clearly, $g \in \Gamma^*$, where $g(x,y) = f(x,y) + f(y,x)$ has the following property: $g(a,a), g(b,b) > g(a,b) = g(b,a)$. Let $\alpha = g(a,a)$ and $\beta = g(b,b)$. If $\alpha \neq \beta$, let $\alpha < \beta$ (the other case is analogous). Using unary cost functions with cost $(\beta - \alpha)/2$, we can construct $h \in \Gamma^*$ satisfying $h(a,a) = h(b,b) > h(a,b) = h(b,a)$. Now if $h(a,a) = h(b,b) = 1$ and $h(a,b) = h(b,a) = 0$, this would correspond to the Max-SAT problem with XOR clauses, which is NP-hard [34]. Since adding a constant to all cost functions and scaling all costs by a constant factor do not affect the difficulty of solving a VCSP instance, and $\Gamma$ is conservative, we can conclude that $\Gamma$ is intractable.

Now suppose that exactly one of the assignments $(a,a),(b,b)$ is in $\mathrm{dom} f$; without loss of generality, let us assume that $(a,a) \in \mathrm{dom} f$ and $(b,b) \notin \mathrm{dom} f$. Thus $g \in \Gamma$, where $g(a,a) = g(a,b) = g(b,a) = 0$ and $g(b,b) = \infty$. Using unary cost functions, we can encode the maximum independent set problem in graphs, a well-known NP-hard problem [35]: every vertex is represented by a node with domain $\{a,b\}$ ($a$ represents not in the set, $b$ represents in the set); an edge between two vertices imposes a term with the cost function $g$ between the corresponding two nodes. For every node, there is a unary term with cost function $h$, where $h(a) = 1$ and $h(b) = 0$. It is clear that minimising the number of nodes assigned $a$ is the same as maximising the number of nodes assigned $b$, thus finding a maximum independent set in the graph. □

### 4.2 Properties of graph $G_\Gamma$

From now on we assume that $E$ does not have soft self-loops. Our goal is to show that $\Gamma$ admits an STP on $M$.

In the lemma below, a *path* of length $k$ is a sequence of edges $\{p_0,p_1\},\{p_1,p_2\},\ldots,\{p_{k-1},p_k\}$, where $\{p_{i-1},p_i\} \in E$. Note that we allow edge repetitions. A path is *even* iff its length is even. A path is a *cycle* if $p_0 = p_k$. If $X \subseteq P$ then $(X, E[X])$ denotes the subgraph of $(P,E)$ induced by $X$.

**Lemma 11.** *Graph $G_\Gamma = (P,E)$ satisfies the following properties:*

*(a) $\{p,q\} \in E$ implies $\{\bar{p},\bar{q}\} \in E$ and vice versa. The two edges are either both soft or both hard.*

*(b) Suppose that $\{p,q\} \in E$ and $\{q,r\} \in E$, then $\{p,\bar{r}\} \in E$. If at least one of the first two edges is soft then the third edge is also soft.*

*(c) For each $p \in P$, nodes $p$ and $\bar{p}$ are either both in $M$ or both in $\overline{M}$.*

*(d) There are no edges from $M$ to $\overline{M}$.*

*(e) Graph $(M, E[M])$ does not have odd cycles.*

*(f) Graph $(M, E[M])$ does not have even paths from $(a,b)$ to $(b,a)$ where $(a,b) \in P$.*

*(g) Nodes $p \in \overline{M}$ do not have incident soft edges.*

*Proof.* **(a)** Follows from the definition.

**(b)** Let $p = (a_1,b_1)$, $q = (a_2,b_2)$ and $r = (a_3,b_3)$. From the definition of the graph, let $f,g \in \Gamma^*$ be binary cost functions such that $(*)$ $f(a_1,a_2)+f(b_1,b_2) > f(a_1,b_2)+f(b_1,a_2)$ and $g(a_2,a_3)+g(b_2,b_3) > g(a_2,b_3) + g(b_2,a_3)$. Without loss of generality, we can assume that

$$\begin{aligned} f(a_1,a_2) &= \alpha, \quad f(a_1,b_2) = f(b_1,a_2) = \gamma, \quad f(b_1,b_2) = \alpha' \\ g(a_2,a_3) &= \beta, \quad g(a_2,b_3) = f(b_2,a_3) = \gamma, \quad g(b_2,b_3) = \beta' \end{aligned} \tag{3}$$

This can be achieved by replacing $f$ with $f'(x,y) = f(x,y) + f(y,x)$ and adding a constant, and similarly for $g$; condition $(*)$ and the complexity of $\Gamma$ are unaffected. From $(*)$ we get $\alpha + \alpha' > 2\gamma$; thus, by adding unary terms to $f$ we can ensure that $\alpha > \gamma$, $\alpha' > \gamma$ and $f(x,y) \geq 0$ for all $x, y \in D$. Similarly, we can assume that $\beta > \gamma$, $\beta' > \gamma$ and $g(x,y) \geq 0$ for all $x, y \in D$.

Let $h(x,z) = \min_{z \in D}\{f(x,y) + u_{\{a_2,b_2\}}(y) + g(y,z)\}$, where $u_{\{a_2,b_2\}}(y) = 0$ if $y \in \{a_2, b_2\}$, and $u_{\{a_2,b_2\}}(y) = C$ otherwise. Here $C$ is a sufficiently large constant, namely $C > \max\{\alpha,\alpha'\} + \max\{\beta,\beta'\}$. From the definition of $h$ and (3) we get $h(a_1,a_3) = h(b_1,b_3) = 2\gamma$. Moreover, $h(a_1,b_3) \geq \min\{\alpha,\beta',C\}$ and $h(b_1,a_3) \geq \min\{\alpha',\beta,C\}$, where $h(a_1,b_3) \geq C$ iff $\alpha = \beta' = \infty$, and $h(b_1,a_3) \geq C$ iff $\alpha' = \beta = \infty$. Therefore, $h(a_1,b_3) + h(b_1,a_3) > h(a_1,a_3) + h(b_1,b_3)$, and so $\{p,\bar{r}\} \in E$.

Now suppose that at least of one of the edges $\{p,q\}$, $\{q,r\}$ is soft, then we can assume that either $(\alpha,\alpha') \neq (\infty,\infty)$ or $(\beta,\beta') \neq (\infty,\infty)$. In each case either $h(a_1,b_3) < C$ or $h(b_1,a_3) < C$, and thus $\{p,\bar{r}\}$ is soft.

**(c)** Follows from (a).

**(d)** Suppose $\{p,q\} \in E$ and $q \in \overline{M}$. The latter fact implies $\{q,q\} \in E$, so by (b) we have $\{p,\bar{q}\} \in E$. From (a) we also get $\{q,\bar{p}\} \in E$. Applying (b) again gives $\{p,p\} \in E$. Thus $p \in \overline{M}$.

**(e)** We prove by induction on $k$ that $(M, E[M])$ does not have cycles of length $2k+1$. For $k=0$ the claim is by assumption (nodes of $M$ do not have self-loops). Suppose it holds for $k \geq 0$, and suppose that $(M, E[M])$ has a cycle $\mathcal{P}, \{p,q\}, \{q,r\}, \{r,s\}$ of length $2k+3$ where $\mathcal{P}$ is path from $s \in M$ to $p \in M$ of length $2k$. Properties (b) and (a) give respectively $\{p,\bar{r}\} \in E$ and $\{\bar{r},\bar{s}\} \in E$. Applying (b) again gives $\{p,s'\} \in E$, therefore $(M, E[M])$ has a cycle $\mathcal{P}, \{p,s\}$ of length $2k+1$. This contradicts to the induction hypothesis.

**(f)** Suppose $(M, E[M])$ has an even path $\mathcal{P}, \{p,q\}, \{q,\bar{r}\}$ where $\mathcal{P}$ is an even path from $r$ to $p$. Property (b) gives $\{p,r\} \in E$, so $(M, E[M])$ has an odd cycle $\mathcal{P}, \{p,r\}$. This contradicts to (e).

**(g)** Suppose $p \in \overline{M}$ (implying $E$ has a hard self-loop $\{p,p\}$) and $\{p,q\}$ is a soft edge in $E$. Properties (b) and (a) give respectively $\{p,\bar{q}\} \in E$ and $\{\bar{q},\bar{p}\} \in E$, and furthermore both edges are soft. Applying (b) again gives that $\{p,p\} \in E$ and this edge is soft. This contradicts to the assumption that $(P,E)$ does not have soft self-loops. ☐

### 4.3 Constructing $\langle \sqcap, \sqcup \rangle$

In this section we complete the proof of Theorem 9 by constructing a pair of operations $\langle \sqcap, \sqcup \rangle$ for $\Gamma$ that behaves as an STP on $M$ and as a multi-projection (returning its two arguments in the same order) on $\overline{M}$.

**Lemma 12.** *There exists an assignment $\sigma : M \to \{-1,+1\}$ such that (i) $\sigma(p) = -\sigma(q)$ for all edges $\{p,q\} \in E$, and (ii) $\sigma(p) = -\sigma(\bar{p})$ for all $p \in M$.*

*Proof.* Let $(M_1,E_1), \ldots, (M_k,E_k)$ be the connected components of graph $(M, E[M])$. Consider the following algorithm for constructing $\sigma$:

S0 Set $\sigma(p) = \varnothing$ ("unassigned") for all $p \in M$.

S1 **for** $i = 1, \ldots, k$

S2 Choose $p_i \in M_i$. If $\sigma(\bar{p}_i)$ is assigned set $\sigma(p_i) = -\sigma(\bar{p}_i)$, otherwise set $\sigma(p_i)$ to $-1$ or $+1$ arbitrarily.

S3 For $p \in M_i - \{p_i\}$ set $\sigma(p) = (-1)^k \sigma(p_i)$ where $k$ is the length of a path from $p_i$ to $p$. (All paths from $p_i$ to $p$ must have the same parity since, by Lemma 11(d), there is no odd cycle in $(M_i, E_i)$.)

S4 **end for**

It can be seen that the algorithm produces an assignment $\sigma$ with claimed properties. Showing property (i) is straightforward, let us prove that (ii) holds. If $p$ and $\bar{p}$ belong to the same connected component, then, by Lemma 11(f), a path from $p$ to $\bar{p}$ must be odd, so $\sigma(p) = -\sigma(\bar{p})$. Suppose that $\bar{p} \in M_j$ and $p \in M_i$ where $j < i$, i.e. $\sigma(\bar{p})$ was assigned first. Let $p_i$ be the element of $M_i$ chosen in step S2. We know that there is a path from $p_i$ to $p$ of some length $k$. Lemma 11(a) implies that there is a path from $\bar{p}$ to $\bar{p}_i$ of the same length $k$. Therefore, $\bar{p}_i \in M_j$, so $\sigma(\bar{p}_i)$ would have already been assigned in step S2 for component $M_i$ and thus we would set $\sigma(p_i) = -\sigma(\bar{p}_i)$. The claim now follows from

$$\sigma(p) = (-1)^k \sigma(p_i) = -(-1)^k \sigma(\bar{p}_i) = -(-1)^k \cdot (-1)^k \sigma(\bar{p}) = -\sigma(\bar{p})$$

□

Given assignment $\sigma$ constructed in Lemma 12, we now define operations $\sqcap, \sqcup : D^2 \to D$ as follows:

- $a \sqcap a = a \sqcup a = a$ for $a \in D$.
- If $(a, b) \in M$ then $a \sqcap b$ and $a \sqcup b$ are the unique elements of $D$ satisfying $\{a \sqcap b, a \sqcup b\} = \{a, b\}$ and $\sigma(a \sqcap b, a \sqcup b) = +1$.
- If $(a, b) \in \overline{M}$ then $a \sqcap b = a$ and $a \sqcup b = b$.

**Lemma 13.** *For any binary cost function $f \in \Gamma^*$ and any $\boldsymbol{x}, \boldsymbol{y} \in \mathtt{dom} f$ there holds*

$$f(\boldsymbol{x} \sqcap \boldsymbol{y}) + f(\boldsymbol{x} \sqcup \boldsymbol{y}) \leq f(\boldsymbol{x}) + f(\boldsymbol{y}) \tag{4}$$

*Proof.* Denote $(a, a') = \boldsymbol{x} \sqcap \boldsymbol{y}$ and $(b, b') = \boldsymbol{x} \sqcup \boldsymbol{y}$. We can assume without loss of generality that $\{\boldsymbol{x}, \boldsymbol{y}\} \neq \{(a, a'), (b, b')\}$, otherwise the claim is straightforward. It is easy to check that the assumption has two implications: (i) $a \neq b$ and $a' \neq b'$; (ii) $\{\boldsymbol{x}, \boldsymbol{y}\} = \{(a, b'), (b, a')\}$.

If $f(a, a') + f(b, b') = f(a, b') + f(b, a')$, then (4) holds trivially. If $f(a, a') + f(b, b') \neq f(a, b') + f(b, a')$, then $E$ contains at least one of the edges $\{(a, b), (a', b')\}$, $\{(a, b), (b', a')\}$. By Lemma 11(c) and Lemma 11(d), pairs $(a, b)$ and $(a', b')$ must either be both in $\overline{M}$ or both in $M$. In the former case (4) is a trivial equality from the definition of $\sqcap$ and $\sqcup$, so we assume the latter case.

The definition of $\sqcap, \sqcup$ and the fact that $(a, a') = \boldsymbol{x} \sqcap \boldsymbol{y}$ and $(b, b') = \boldsymbol{x} \sqcup \boldsymbol{y}$ imply that $\sigma(a, b) = \sigma(a', b') = +1$. Thus, set $E$ does not have edge $((a, b), (a', b'))$, and therefore

$$f(a, a') + f(b, b') \leq f(a, b') + f(b, a')$$

which is equivalent to (4). □

In order to proceed, we introduce the following notation. Given a cost function $f$ of arity $m$, we denote by $V$ the set of variables corresponding to the arguments of $f$, with $|V| = m$. For two assignments $\boldsymbol{x}, \boldsymbol{y} \in D^m$ we denote $\Delta(\boldsymbol{x}, \boldsymbol{y}) = \{i \in V \mid x_i \neq y_i\}$ to be the set of variables on which $\boldsymbol{x}$ and $\boldsymbol{y}$ differ.

**Lemma 14.** *Condition* (4) *holds for any cost function $f \in \Gamma^*$ and assignments $\boldsymbol{x}, \boldsymbol{y} \in \mathtt{dom} f$ with $|\Delta(\boldsymbol{x}, \boldsymbol{y})| \leq 2$.*

*Proof.* If $|\Delta(\boldsymbol{x}, \boldsymbol{y})| \leq 1$ then $\{\boldsymbol{x} \sqcap \boldsymbol{y}, \boldsymbol{x} \sqcup \boldsymbol{y}\} = \{\boldsymbol{x}, \boldsymbol{y}\}$, so the claim is trivial. We now prove it in the case $|\Delta(\boldsymbol{x}, \boldsymbol{y})| = 2$ using induction on $|V|$. The base case $|V| = 2$ follows from Lemma 13; suppose that $|V| \geq 3$. Choose $k \in V - \Delta(\boldsymbol{x}, \boldsymbol{y})$. For simplicity of notation, let us assume that $k$ corresponds to the first argument of $f$. Define cost function of $|V| - 1$ variables as

$$g(\boldsymbol{z}) = \min_{a \in D} \{u(a) + f(a, \boldsymbol{z})\} \qquad \forall \boldsymbol{z} \in D^{V - \{k\}} \tag{5}$$

where $u$ is the following unary cost function: $u(a) = 0$ if $a = x_k = y_k$, and $u(a) = C$ otherwise. Here $C$ is a sufficiently large constant, namely $C > f(\boldsymbol{x}) + f(\boldsymbol{y})$.

Let $\hat{\boldsymbol{x}}$ and $\hat{\boldsymbol{y}}$ be the restrictions of respectively $\boldsymbol{x}$ and $\boldsymbol{y}$ to $V - \{k\}$. Clearly, $g \in \Gamma^*$, $g(\hat{\boldsymbol{x}}) = f(\boldsymbol{x}) < \infty$ and $g(\hat{\boldsymbol{y}}) = f(\boldsymbol{y}) < \infty$. By the induction hypothesis

$$g(\hat{\boldsymbol{x}} \sqcap \hat{\boldsymbol{y}}) + g(\hat{\boldsymbol{x}} \sqcup \hat{\boldsymbol{y}}) \leq g(\hat{\boldsymbol{x}}) + g(\hat{\boldsymbol{y}}) = f(\boldsymbol{x}) + f(\boldsymbol{y}) \tag{6}$$

This implies that $g(\hat{\boldsymbol{x}} \sqcap \hat{\boldsymbol{y}}) < C$, which is possible only if $g(\hat{\boldsymbol{x}} \sqcap \hat{\boldsymbol{y}}) = f(a, \hat{\boldsymbol{x}} \sqcap \hat{\boldsymbol{y}}) = f(\boldsymbol{x} \sqcap \boldsymbol{y})$ where $a = x_k = y_k$. Similarly, $g(\hat{\boldsymbol{x}} \sqcup \hat{\boldsymbol{y}}) = f(a, \hat{\boldsymbol{x}} \sqcup \hat{\boldsymbol{y}}) = f(\boldsymbol{x} \sqcup \boldsymbol{y})$. Thus, (6) is equivalent to (4). ☐

**Lemma 15.** *Condition (4) holds for any cost function $f \in \Gamma^*$ and any $\boldsymbol{x}, \boldsymbol{y} \in \texttt{dom} f$.*

*Proof.* We use induction on $|\Delta(\boldsymbol{x}, \boldsymbol{y})|$. The base case $|\Delta(\boldsymbol{x}, \boldsymbol{y})| \leq 2$ follows from Lemma 14; suppose that $|\Delta(\boldsymbol{x}, \boldsymbol{y})| \geq 3$. Let us partition $\Delta(\boldsymbol{x}, \boldsymbol{y})$ into three sets $A, B, C$ as follows:

$$\begin{aligned} A &= \{i \in \Delta(\boldsymbol{x}, \boldsymbol{y}) \mid (x_i, y_i) \in M, \;\; x_i = x_i \sqcap y_i, \;\; y_i = x_i \sqcup y_i\} \\ B &= \{i \in \Delta(\boldsymbol{x}, \boldsymbol{y}) \mid (x_i, y_i) \in M, \;\; x_i = x_i \sqcup y_i, \;\; y_i = x_i \sqcap y_i\} \\ C &= \{i \in \Delta(\boldsymbol{x}, \boldsymbol{y}) \mid (x_i, y_i) \in \overline{M}\} \end{aligned}$$

Two cases are possible.

<u>**Case 1**</u> $|A \cup C| \geq 2$. Let us choose variable $k \in A \cup C$, and define assignments $\boldsymbol{x}', \boldsymbol{y}'$ as follows: $x'_i = y'_i = x_i = y_i$ if $x_i = y_i$, and for other variables

$$x'_i = \begin{cases} x_i & \text{if } i = k \\ y_i & \text{if } i \in (A \cup C) - \{k\} \\ x_i & \text{if } i \in B \end{cases} \qquad y'_i = \begin{cases} x_i & \text{if } i = k \\ y_i & \text{if } i \in (A \cup C) - \{k\} \\ y_i & \text{if } i \in B \end{cases}$$

It can be checked that

$$\boldsymbol{x} \sqcap \boldsymbol{y}' = \boldsymbol{x} \sqcap \boldsymbol{y} \qquad \boldsymbol{x} \sqcup \boldsymbol{y}' = \boldsymbol{x}' \qquad \boldsymbol{x}' \sqcap \boldsymbol{y} = \boldsymbol{y}' \qquad \boldsymbol{x}' \sqcup \boldsymbol{y} = \boldsymbol{x} \sqcup \boldsymbol{y}$$

Furthermore, $\Delta(\boldsymbol{x}, \boldsymbol{y}') = \Delta(\boldsymbol{x}, \boldsymbol{y}) - \{k\}$ and $\Delta(\boldsymbol{x}', \boldsymbol{y}) = \Delta(\boldsymbol{x}, \boldsymbol{y}) - ((A \cup C) - \{k\})$ so by the induction hypothesis

$$f(\boldsymbol{x} \sqcap \boldsymbol{y}) + f(\boldsymbol{x}') \leq f(\boldsymbol{x}) + f(\boldsymbol{y}') \tag{7}$$

assuming that $\boldsymbol{y}' \in \texttt{dom} f$, and

$$f(\boldsymbol{y}') + f(\boldsymbol{x} \sqcup \boldsymbol{y}) \leq f(\boldsymbol{x}') + f(\boldsymbol{y}) \tag{8}$$

assuming that $\boldsymbol{x}' \in \texttt{dom} f$. Two cases are possible:

- $\boldsymbol{y}' \in \texttt{dom} f$. Inequality (7) implies that $\boldsymbol{x}' \in \texttt{dom} f$. The claim then follows from summing (7) and (8).

- $\boldsymbol{y}' \notin \texttt{dom} f$. Inequality (8) implies that $\boldsymbol{x}' \notin \texttt{dom} f$. Assume for simplicity of notation that $k$ corresponds to the first argument of $f$. Define cost function of $|V| - 1$ variables

  $$g(\boldsymbol{z}) = \min_{a \in D} \{u(a) + f(a, \boldsymbol{z})\} \qquad \forall \boldsymbol{z} \in D^{V - \{k\}}$$

  where $u(a)$ is the following unary cost function: $u(x_k) = 0$, $u(y_k) = C$ and $u(a) = 2C$ for $a \in D - \{x_k, y_k\}$. Here $C$ is a sufficiently large constant, namely $C > f(\boldsymbol{x}) + f(\boldsymbol{y})$.

Let $\hat{\boldsymbol{x}}, \hat{\boldsymbol{y}}, \hat{\boldsymbol{x}}', \hat{\boldsymbol{y}}'$ be restrictions of respectively $\boldsymbol{x}, \boldsymbol{y}, \boldsymbol{x}', \boldsymbol{y}'$ to $V - \{k\}$. Clearly, $g \in \Gamma^*$ and

$$\begin{aligned} g(\hat{\boldsymbol{y}}) = g(\hat{\boldsymbol{y}}') &= u(y_k) + f(y_k, \hat{\boldsymbol{y}}) = f(\boldsymbol{y}) + C \qquad &\text{(since } (x_k, \hat{\boldsymbol{y}}) = \boldsymbol{y}' \notin \texttt{dom} f) \\ g(\hat{\boldsymbol{x}}) &= f(x_k, \hat{\boldsymbol{x}}) = f(\boldsymbol{x}) \end{aligned}$$

By the induction hypothesis

$$g(\hat{\boldsymbol{x}} \sqcap \hat{\boldsymbol{y}}) + g(\hat{\boldsymbol{x}} \sqcup \hat{\boldsymbol{y}}) \leq g(\hat{\boldsymbol{x}}) + g(\hat{\boldsymbol{y}}) = f(\boldsymbol{x}) + f(\boldsymbol{y}) + C \tag{9}$$

We have $g(\hat{\boldsymbol{x}} \sqcup \hat{\boldsymbol{y}}) < 2C$, so we must have either $g(\hat{\boldsymbol{x}} \sqcup \hat{\boldsymbol{y}}) = f(x_k, \hat{\boldsymbol{x}} \sqcup \hat{\boldsymbol{y}})$ or $g(\hat{\boldsymbol{x}} \sqcup \hat{\boldsymbol{y}}) = f(y_k, \hat{\boldsymbol{x}} \sqcup \hat{\boldsymbol{y}}) + C = f(\boldsymbol{x} \sqcup \boldsymbol{y}) + C$. The former case is impossible since $(x_k, \hat{\boldsymbol{x}} \sqcup \hat{\boldsymbol{y}}) = \boldsymbol{x}' \notin \texttt{dom} f$, so $g(\hat{\boldsymbol{x}} \sqcup \hat{\boldsymbol{y}}) = f(\boldsymbol{x} \sqcup \boldsymbol{y}) + C$. Combining it with (9) gives

$$g(\hat{\boldsymbol{x}} \sqcap \hat{\boldsymbol{y}}) + f(\boldsymbol{x} \sqcup \boldsymbol{y}) \leq f(\boldsymbol{x}) + f(\boldsymbol{y}) \tag{10}$$

This implies that $g(\hat{\boldsymbol{x}} \sqcap \hat{\boldsymbol{y}}) < C$, so we must have $g(\hat{\boldsymbol{x}} \sqcap \hat{\boldsymbol{y}}) = f(x_k, \hat{\boldsymbol{x}} \sqcap \hat{\boldsymbol{y}}) = f(\boldsymbol{x} \sqcap \boldsymbol{y})$. Thus, (10) is equivalent to (4).

**Case 2** $|B| \geq 2$. Let us choose variable $k \in B$, and define assignments $\boldsymbol{x}', \boldsymbol{y}'$ as follows: $x'_i = y'_i = x_i = y_i$ if $x_i = y_i$, and for other variables

$$x'_i = \begin{cases} y_i & \text{if } i = k \\ x_i & \text{if } i \in A \cup C \\ x_i & \text{if } i \in B - \{k\} \end{cases} \qquad y'_i = \begin{cases} y_i & \text{if } i = k \\ y_i & \text{if } i \in A \cup C \\ x_i & \text{if } i \in B - \{k\} \end{cases}$$

It can be checked that

$$\boldsymbol{x}' \sqcap \boldsymbol{y} = \boldsymbol{x} \sqcap \boldsymbol{y} \qquad \boldsymbol{x}' \sqcup \boldsymbol{y} = \boldsymbol{y}' \qquad \boldsymbol{x} \sqcap \boldsymbol{y}' = \boldsymbol{x}' \qquad \boldsymbol{x} \sqcup \boldsymbol{y}' = \boldsymbol{x} \sqcup \boldsymbol{y}$$

Furthermore, $\Delta(\boldsymbol{x}', \boldsymbol{y}) = \Delta(\boldsymbol{x}, \boldsymbol{y}) - \{k\}$ and $\Delta(\boldsymbol{x}, \boldsymbol{y}') = \Delta(\boldsymbol{x}, \boldsymbol{y}) - (B - \{k\})$ so by the induction hypothesis

$$f(\boldsymbol{x} \sqcap \boldsymbol{y}) + f(\boldsymbol{y}') \leq f(\boldsymbol{x}') + f(\boldsymbol{y}) \tag{11}$$

assuming that $\boldsymbol{x}' \in \texttt{dom} f$, and

$$f(\boldsymbol{x}') + f(\boldsymbol{x} \sqcup \boldsymbol{y}) \leq f(\boldsymbol{x}) + f(\boldsymbol{y}') \tag{12}$$

assuming that $\boldsymbol{y}' \in \texttt{dom} f$. Two cases are possible:

- $\boldsymbol{x}' \in \texttt{dom} f$. Inequality (11) implies that $\boldsymbol{y}' \in \texttt{dom} f$. The claim then follows from summing (11) and (12).
- $\boldsymbol{x}' \notin \texttt{dom} f$. Inequality (12) implies that $\boldsymbol{y}' \notin \texttt{dom} f$. Assume for simplicity of notation that $k$ corresponds to the first argument of $f$. Define function of $|V| - 1$ variables

$$g(\boldsymbol{z}) = \min_{a \in D} \{u(a) + f(a, \boldsymbol{z})\} \qquad \forall \boldsymbol{z} \in D^{V - \{k\}}$$

where $u(a)$ is the following unary term: $u(y_k) = 0$, $u(x_k) = C$ and $u(a) = 2C$ for $a \in D - \{x_k, y_k\}$. Here $C$ is a sufficiently large constant, namely $C > f(\boldsymbol{x}) + f(\boldsymbol{y})$.

Let $\hat{\boldsymbol{x}}, \hat{\boldsymbol{y}}, \hat{\boldsymbol{x}}', \hat{\boldsymbol{y}}'$ be restrictions of respectively $\boldsymbol{x}, \boldsymbol{y}, \boldsymbol{x}', \boldsymbol{y}'$ to $V - \{k\}$. Clearly, $g \in \Gamma^*$ and

$$\begin{aligned} g(\hat{\boldsymbol{x}}) = g(\hat{\boldsymbol{x}}') &= u(x_k) + f(x_k, \hat{\boldsymbol{x}}) = f(\boldsymbol{x}) + C \qquad &\text{(since } (y_k, \hat{\boldsymbol{x}}) = \boldsymbol{x}' \notin \texttt{dom} f) \\ g(\hat{\boldsymbol{y}}) &= f(y_k, \hat{\boldsymbol{y}}) = f(\boldsymbol{y}) \end{aligned}$$

By the induction hypothesis

$$g(\hat{\boldsymbol{x}} \sqcap \hat{\boldsymbol{y}}) + g(\hat{\boldsymbol{x}} \sqcup \hat{\boldsymbol{y}}) \le g(\hat{\boldsymbol{x}}) + g(\hat{\boldsymbol{y}}) = f(\boldsymbol{x}) + f(\boldsymbol{y}) + C \tag{13}$$

We have $g(\hat{\boldsymbol{x}} \sqcup \hat{\boldsymbol{y}}) < 2C$, so we must have either $g(\hat{\boldsymbol{x}} \sqcup \hat{\boldsymbol{y}}) = f(y_k, \hat{\boldsymbol{x}} \sqcup \hat{\boldsymbol{y}})$ or $g(\hat{\boldsymbol{x}} \sqcup \hat{\boldsymbol{y}}) = f(x_k, \hat{\boldsymbol{x}} \sqcup \hat{\boldsymbol{y}}) + C = f(\boldsymbol{x} \sqcup \boldsymbol{y}) + C$. The former case is impossible since $(y_k, \hat{\boldsymbol{x}} \sqcup \hat{\boldsymbol{y}}) = \boldsymbol{y}' \notin \mathtt{dom} f$, so $g(\hat{\boldsymbol{x}} \sqcup \hat{\boldsymbol{y}}) = f(\boldsymbol{x} \sqcup \boldsymbol{y}) + C$. Combining it with (13) gives

$$g(\hat{\boldsymbol{x}} \sqcap \hat{\boldsymbol{y}}) + f(\boldsymbol{x} \sqcup \boldsymbol{y}) \le f(\boldsymbol{x}) + f(\boldsymbol{y}) \tag{14}$$

This implies that $g(\hat{\boldsymbol{x}} \sqcap \hat{\boldsymbol{y}}) < C$, so we must have $g(\hat{\boldsymbol{x}} \sqcap \hat{\boldsymbol{y}}) = f(y_k, \hat{\boldsymbol{x}} \sqcap \hat{\boldsymbol{y}}) = f(\boldsymbol{x} \sqcap \boldsymbol{y})$. Thus, (14) is equivalent to (4).

□

# 5 Discussion of the general case

While the case of conservative finite-valued languages has now been completely characterized, the case of general conservative languages (i.e. languages that allow both finite and infinite costs and contain all finite-valued unary functions) remains open. We conjecture that the tractability criterion is the same as for finite-valued languages: $\Gamma$ is tractable iff $G_\Gamma$ does not contain a soft self-loop. The "only if" part is proven in Theorem 9(a). It remains to show that if $G_\Gamma$ does not contain a soft self-loop, then $\Gamma$ is tractable. We know that in that case $\Gamma$ admits a multimorphism, constructed in Section 4.3, that behaves as an STP on $M$.

Below we review some known partial results related to the general case. These results use notions of *polymorphisms* and *multimorphisms* that generalise Definition 3.

**Definition 16.** *A mapping $F : D^k \to D$, $k \ge 1$ is called a* polymorphism *of a cost function $f : D^m \to \overline{\mathbb{R}}_+$ if*

$$F(\boldsymbol{x}_1, \ldots, \boldsymbol{x}_k) \in \mathtt{dom} f \qquad \forall \boldsymbol{x}_1, \ldots, \boldsymbol{x}_k \in \mathtt{dom} f$$

*where $F$ is applied component-wise. $F$ is a polymorphism of a language $\Gamma$ if $F$ is a polymorphism of every cost function in $\Gamma$.*

**Definition 17** ([22])**.** *A mapping $\langle F_1, \ldots, F_k \rangle : D^k \to D^k$, where $F_i : D^k \to D$, $1 \le i \le k$, is a* multimorphism *of a cost function $f : D^m \to \overline{\mathbb{R}}_+$ if*

$$\sum_{i=1}^{k} f(F_i(\boldsymbol{x}_1, \ldots, \boldsymbol{x}_k)) \le \sum_{i=1}^{k} f(\boldsymbol{x}_i) \qquad \forall \boldsymbol{x}_1, \ldots, \boldsymbol{x}_k \in \mathtt{dom} f$$

*where $F_i$ are applied coordinate-wise. $\langle F_1, \ldots, F_k \rangle$ is a multimorphism of a language $\Gamma$ if $\langle F_1, \ldots, F_k \rangle$ is a multimorphism of every cost function in $\Gamma$.*

An STP, defined in Definition 3, is an example of a binary multimorphism. Note, if $\langle F_1, \ldots, F_k \rangle$ is a multimorphism of $\Gamma$, then each mapping $F_i$, $1 \le i \le k$ is a polymorphism of $\Gamma$.

## 5.1 Boolean case ($|D| = 2$)

This case has been completely characterized by Cohen *et al.* [22].

**Theorem 18** ([22]). *Let $\Gamma$ be a Boolean conservative language.*

1. *Suppose $\Gamma$ is finite-valued. If $\Gamma$ admits an STP multimorphism then it is tractable, otherwise it is NP-hard.*
2. *Suppose $\Gamma$ is general-valued. If $\Gamma$ admits of the following as a multimorphism then it is tractable:*
   - *STP multimorphism*
   - *$\langle \mathtt{Mj}, \mathtt{Mj}, \mathtt{Mn} \rangle$ multimorphism where $\mathtt{Mj}$ and $\mathtt{Mn}$ are the unique* majority *and* minority *operations defined by $\mathtt{Mj}(a,a,b) = \mathtt{Mj}(a,b,a) = \mathtt{Mj}(b,a,a) = a$, $\mathtt{Mn}(a,a,b) = \mathtt{Mn}(a,b,a) = \mathtt{Mn}(b,a,a) = b$.*

   *Otherwise $\Gamma$ is NP-hard.*

## 5.2 Non-Boolean case ($|D| > 2$)

Dichotomy results have been established for some special cases by Bulatov [12] and Takhanov [27].

**Theorem 19** ([12]). *Let $\Gamma$ be a crisp language containing all possible $\{0, \infty\}$-valued unary cost functions. Suppose that for every subset $B \subseteq D$ with $|B| = 2$ there exists polymorphism $F^B$ of $\Gamma$ whose restriction $F^B|_B$ to $B$ satisfies one of the following conditions:*

- *$F^B|_B$ is a binary semilattice operation.*
- *$F^B|_B$ is the ternary majority operation.*
- *$F^B|_B$ is the ternary minority operation.*

*Then $\Gamma$ is tractable; otherwise $\Gamma$ is NP-hard.*

**Theorem 20** ([27]). *Let $\Gamma$ be a language containing some crisp functions and additionally all possible finite-valued unary functions, and let $M$ be a subset of $P$ with the property $(a,b)$ and $(b,a)$ are either both in $M$ or both not in $M$. A ternary operation $m : D^3 \to D$ is called* arithmetical *on $\overline{M} = P - M$ if $m(a,b,c) \in \{a,b,c\}$ for all $a,b,c \in D$ and $m(a,a,b) = m(b,a,a) = m(b,a,b) = b$ for all $(a,b) \in \overline{M}$*

*If $\Gamma$ admits an STP multimorphism on $M$ and an arithmetical polymorphism on $\overline{M}$ then $\Gamma$ is tractable. Otherwise $\Gamma$ is NP-hard.*

One can show that the absence of soft self-loops in $G_\Gamma$ implies Takhanov's "necessary local conditions" for the set of polymorphisms of $\Gamma$ (details are omitted). Takhanov's results then imply the existence of an arithmetical polymorphism on $\overline{M}$.

Finally, we mention another special case of tractable conservative languages.

**Theorem 21** ([22]). *Let $\mathtt{Mj}_1$, $\mathtt{Mj}_2$, and $\mathtt{Mn}_3$ be the ternary functions defined on $D$ as follows:*

$$\mathtt{Mj}_1(x,y,z) = \begin{cases} y & \text{if } y = z \\ x & \text{otherwise} \end{cases} \qquad \mathtt{Mj}_2(x,y,z) = \begin{cases} x & \text{if } x = z \\ y & \text{otherwise} \end{cases}$$

$$\mathtt{Mn}_3(x,y,z) = \begin{cases} x & \text{if } y = z \text{ and } z \neq x \\ y & \text{if } x = z \text{ and } z \neq y \\ z & \text{otherwise} \end{cases}$$

*If $\Gamma$ admits multimorphism $\langle \mathtt{Mj}_1, \mathtt{Mj}_2, \mathtt{Mj}_3 \rangle$ then $\Gamma$ is tractable.*